\documentclass[letterpaper]{article} 
\usepackage{aaai2026}  
\usepackage{times}  
\usepackage{helvet}  
\usepackage{courier}  
\usepackage[hyphens]{url}  
\usepackage{graphicx} 
\usepackage{natbib}  
\usepackage{caption} 
\usepackage{algorithm}
\usepackage{algorithmic}

\usepackage{amsmath}
\usepackage{multirow}
\usepackage[most]{tcolorbox}
\usepackage[dvipsnames]{xcolor}
\usepackage{enumitem}
\definecolor{lightorange}{RGB}{255,200,100} 
\definecolor{lightblue}{RGB}{173,216,230} 
\usepackage{listings}
\DeclareCaptionStyle{ruled}{labelfont=normalfont,labelsep=colon,strut=off} 
\title{An LLM-based Quantitative Framework for\\Evaluating High-Stealthy Backdoor Risks in OSS Supply Chains}
\author {
	Zihe Yan\textsuperscript{\rm 1}\thanks{These authors contributed equally. $\dagger$Corresponding authors. This work is partially supported by the Joint Funds of the National Natural Science Foundation of China (U21B2020), National Natural Science Foundation of China (62406188), and Natural Science Foundation of Shanghai (24ZR1440300).}
	Kai Luo\textsuperscript{\rm 2}\footnotemark[1],
	Haoyu Yang\textsuperscript{\rm 3},
	Yang Yu\textsuperscript{\rm 3},
	Zhuosheng Zhang\textsuperscript{\rm 1}\footnotemark[2],
	Guancheng Li\textsuperscript{\rm 3}\footnotemark[2]
}
\affiliations {
	\textsuperscript{\rm 1}School of Computer Science, Shanghai Jiao Tong University, China\\
	\textsuperscript{\rm 2}Institute for Interdisciplinary Information Sciences, Tsinghua University, China\\
	\textsuperscript{\rm 3}Tencent Xuanwu Lab, Tencent, China\\
	\texttt{\{yangtuomao,zhangzs\}@sjtu.edu.cn} \\
	\texttt{luok22@mails.tsinghua.edu.cn}\\
	\texttt{\{haoyuyang,tkyu,atumli\}@tencent.com}
}

\usepackage{bibentry}
\begin{document}
	
	\maketitle
	
	\begin{abstract}
		In modern software development workflows, the open-source software supply chain significantly contributes to efficient and convenient engineering practices. With increasing system complexity, it has become a common practice to use open-source software as third-party dependencies. However, due to the lack of maintenance for underlying dependencies and insufficient community auditing, ensuring the security of source code and the legitimacy of repository maintainers has become a challenge, particularly in the context of high-stealth backdoor attacks such as the XZ-Util incident. To address these problems, we propose a fine-grained project evaluation framework for backdoor risk assessment in open-source software. Our evaluation framework models highly stealthy backdoor attacks from the attacker’s perspective and defines targeted metrics for each attack stage. Moreover, to overcome the limitations of static analysis in assessing the reliability of repository maintenance activities, such as irregular committer privilege escalation and insufficient review participation, we employ large language models (LLMs) to perform semantic evaluation of code repositories while avoiding reliance on manually crafted patterns. The effectiveness of our framework is validated on 66 high-priority packages in the Debian ecosystem, and the experimental results reveal that the current open-source software supply chain is exposed to a series of security risks.
	\end{abstract}
	
	\begin{links}
		\link{Code}{https://github.com/XuanwuLab/HSBRiskEvaluator}
	\end{links}

	\section{Introduction}
	As the cornerstone of modern software engineering, open-source software has been deeply integrated into a wide range of systems, from desktop applications~\cite{kamau2025investigating} and mobile platforms~\cite{haakegaard2024performance} to cloud computing infrastructures. By leveraging collaborative development, code reuse, and community-driven models, open-source software significantly lowers technical barriers and accelerates software innovation and iteration. Developers worldwide can share components and reuse modules to efficiently build complex application ecosystems. Many mainstream architectures, such as Linux, Docker, and AI frameworks, heavily rely on the openness and reusability of open-source software.
	
	However, the openness and extensive dependency propagation of open-source software have also introduced increasing risks to the software supply chain. In recent years, attackers have gradually evolved traditional poisoning attacks into more targeted and persistent advanced threats. A typical example is the \textbf{XZ Utils incident}~\cite{vayrynen2025human}, where the adversary did not merely rely on a one-off code injection but instead maintained a long-term presence within the community, steadily building trust and waiting for the right moment to strike. These attacks exhibit distinctive characteristics: 
	\begin{itemize}
		\item \textbf{High stealth}, achieved by leveraging techniques such as binary payload embedding~\cite{lin2024hpdh}, formatting obfuscation, and pseudo-random naming to deeply bury backdoors within critical dependency paths;
		\item \textbf{High complexity}, as the attack chains often span multiple repositories and downstream build systems~\cite{shao2024exploring}, making detection through standard review processes extremely difficult;
		\item \textbf{Human-oriented tactics}~\cite{fu2025human}, where attackers use social engineering to integrate into maintainer circles and masquerade as trusted contributors, significantly lowering defenders’ vigilance against potential threats.
	\end{itemize}
	
	As open-source dependencies become increasingly intertwined and automated build tools proliferate, once such backdoors are merged into the main branch, their impact can rapidly propagate across projects and platforms, creating systemic risks that are extremely hard to contain.

	To address these threats, existing studies have explored defense mechanisms based on static code analysis, security auditing, software bill of materials (SBOM) generation~\cite{xia2023empirical}, and dependency tracking~\cite{williams2025research}. However, these approaches often suffer from two limitations: (1) the lack of attacker-centric modeling to anticipate which repositories are likely targets; and (2) the reliance on known patterns, which fails to detect evolving stealth strategies. Consequently, there is an urgent need for a systematic and quantifiable risk assessment method to help developers and platform maintainers identify \textit{high-value poisoning targets} and formulate effective defense strategies.
	
	In this work, we propose a systematic \textbf{risk assessment framework for highly stealthy poisoning attacks} targeting open-source repositories. From the attacker's perspective, our framework models the target selection process based on four core dimensions: (1) \textbf{Dependency Impact}, which measures the propagation effect of the component in the software ecosystem; (2) \textbf{Payload Concealment}, which assesses the feasibility of embedding and disguising malicious code in the project; (3) \textbf{Community Quality}, which captures the strength of community mechanisms in defending against malicious code contributions from attackers; and (4) \textbf{Continuous Integration}, 
	Based on this metric system, we integrate LLMs to perform static and semantic analysis of repositories, ultimately building an automated assessment tool. This tool generates a multi-dimensional \textit{risk matrix} to provide structured, quantitative scores of poisoning risk in open-source software.
	
	The main contributions of this work are as follows:
	\begin{enumerate}
		\item A unified evaluation framework is proposed to quantify and assess supply chain poisoning risks in open-source Linux packages hosted on GitHub. The framework integrates both structural and behavioral risk factors.
		
		\item A fully automated assessment tool is developed and released. The tool leverages both static heuristics and LLM-based semantic analysis to support large-scale, code-aware risk evaluation.
		
		\item We empirically evaluate 66 high-priority packages in the Debian ecosystem, revealing that \textbf{Community Quality} remains the most prevalent attack surface across projects. This insight is further validated through a case study on the \texttt{xz} repository, a real-world incident that reflects our framework’s key risk indicators.
	\end{enumerate}
	\begin{figure*}[th]
		\centering
		\includegraphics[width=0.8\linewidth]{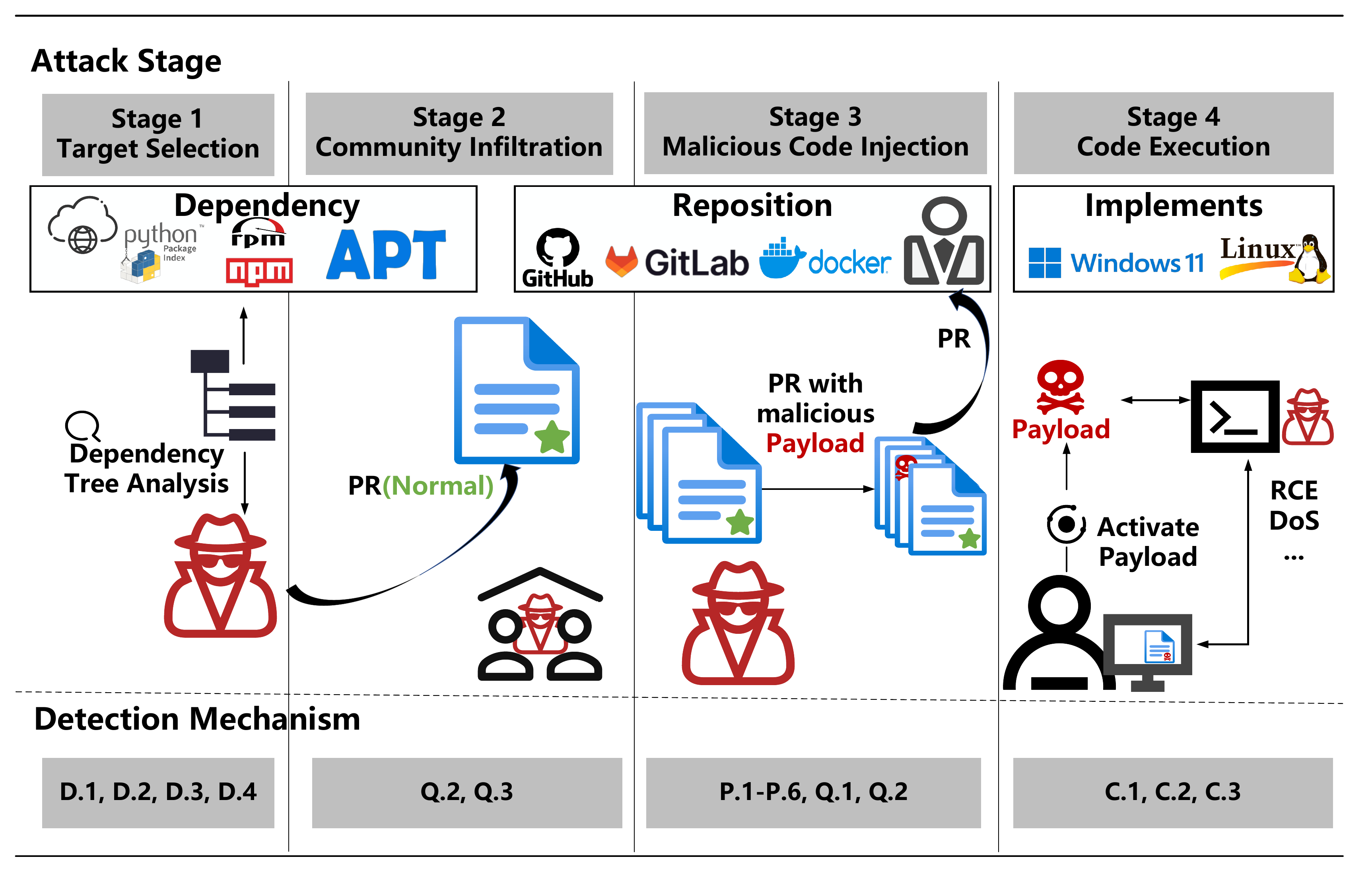}
		\caption{The attack framework and detection mechanism.}
		\label{fig:att_frame}
	\end{figure*}

	\section{Related Work}
	\subsection{Software Supply Chain Attack} 
	Although most work has focused on framework design for efficiency improvement~\cite{vazquez2024blockchain}, security issues have recently attracted increasing attention due to the rising trend of software supply chain attack incidents. 
	Such vulnerability can exist throughout the entire software development lifecycle and pose threats to various dependency environments (e.g., npm, PyPI, GitHub).~\citet{williams2025research} studied how attackers exploit package managers to achieve arbitrary code execution in open-source software supply chains. Their study analyzed seven ecosystems, including npm and PyPI, and identified three install-time and four runtime techniques that attackers use to execute malicious code.~\citet{delicheh2024mitigating} systematically analyzed the software supply chain security issues in GitHub Actions, identifying risks such as dependency confusion, poor permission management, command injection, and credential exposure in reusable Actions and workflows. Because such attacks are often \textbf{stealthily} embedded in dependencies or build processes, they can impact a wide range of downstream projects and users, resulting in extensive and severe consequences.
	
	\subsection{Backdoor Detection and Risk Assessment in Open Source Repositories}
	To systematize the detection of risks in open source software (OSS)~\cite{aberdour2007achieving}, numerous risk assessment frameworks have been proposed, including OpenSSF Scorecard~\cite{zahan2023openssf}, NIST 800-161~\cite{barabanov2020systematics}, and BSIMM~\cite{mcgraw2015software}. These frameworks support both pre-release warnings and post-release issue detection~\cite{barabanov2020systematics}. Building on them, \citet{hamer2025closing} mapped the metrics of ten supply chain frameworks to ATT\&CK attack techniques~\cite{yousaf2024sinking} and analyzed incidents such as SolarWinds~\cite{anisa2024solarwinds}, Log4j~\cite{martinez2021software}, and XZ Utils~\cite{przymus2025wolves}. However, their approach only identified attack-related tasks and could not serve as a general standard for evaluating OSS security. It also overlooked highly stealthy backdoor threats, particularly when attackers infiltrate project communities. To mitigate this gap, \citet{tan2025advanced} introduced human factors as an attack vector, while \citet{kalu2025arms} further proposed reputation-based metrics to assess maintainer trustworthiness. Nevertheless, such reputation scores remain vulnerable to manipulation, as demonstrated by the XZ Utils incident, which undermines their reliability.

	\section{Method}
	\label{sec:method}
	\subsection{Threat Modeling and Attack Framework}
	
	We model the poisoning threat from the perspective of a stealthy and persistent adversary, inspired by real-world incidents such as the \texttt{xz} backdoor. The attacker’s primary objective is to inject malicious payloads into upstream open-source repositories in a way that maximizes downstream propagation and minimizes detection. To support this goal, we define both the adversary’s capabilities and a four-stage attack framework that drives our risk evaluation design.

	\paragraph{Adversary Assumptions}
	The adversary is assumed to possess full access to publicly available metadata of open-source repositories, including dependency trees, version histories, CI configurations, and contribution patterns. The attacker does not control downstream users but can continuously observe ecosystem-wide trends.
	
	Meanwhile, the adversary is capable of submitting well-crafted pull requests (PRs), building trust through benign contributions, and crafting hidden payloads that evade conventional static analysis. We further assume the use of advanced language models to assist in semantic manipulation and stealth injection. However, the attacker does not have knowledge of specific security review policies used by downstream integrators.

	\paragraph{The Attack Framework}

	As illustrated in Figure~\ref{fig:att_frame}, the poisoning process consists of four stages reflecting core structural weaknesses.
	
	\begin{itemize}
		\item \textbf{Stage 1: Reconnaissance and Target Selection.} The attacker analyzes dependency graphs to identify upstream components with high transitive influence.
		\item \textbf{Stage 2: Initial Access via Social Infiltration.} The attacker joins the community by submitting low-risk PRs and builds trust.
		\item \textbf{Stage 3: Hidden Payload Injection.} Carefully crafted triggers are embedded into accepted PRs or auxiliary files, often across multiple repositories.
		\item \textbf{Stage 4: Downstream Execution and Exploitation.} The malicious code is activated during build or runtime in downstream environments.
	\end{itemize}
	
	Each of these stages introduces unique risks, which are systematically captured in our framework via four corresponding dimensions: Dependency Impact (DI), Payload Concealment (PC), Community Quality (CQ), and Continuous Integration (CI). These dimensions form the core of our automated risk evaluation method.
	
	
	\begin{table*}[htbp]
		\centering
		\caption{Structure of the Risk Evaluation Framework and Its Mapping to Attack Stages}
		\label{tab:dimension-score-method}
		\resizebox{\textwidth}{!}{%
			\begin{tabular}{|l|c|c|c|c|c|}
				\hline
				\textbf{Dimension} & \textbf{Index} & \textbf{Individual Risk Metric} & \textbf{Score} & \textbf{Analysis Method} & \textbf{Attack Stage} \\
				\hline
				\multirow{4}{*}{Dependency Impact} 
				& D.1 & Self Priority Exposure & Normalized & Static & S.1 \\
				& D.2 & Self Essential Exposure & Normalized & Static & S.1 \\
				& D.3 & Dependency Priority Exposure & Normalized & Static & S.1 \\
				& D.4 & Dependency Essential Exposure & Normalized & Static & S.1 \\
				\hline
				
				\multirow{6}{*}{Payload Concealment} 
				& P.1 & Binary in Test Files & Boolean & Static and Semantic & S.3 \\
				& P.2 & Binary in Documentation & Boolean & Static and Semantic & S.3 \\
				& P.3 & Binary in Code Files & Boolean & Static and Semantic & S.3 \\
				& P.4 & Binary in Asset Files & Boolean & Static and Semantic & S.3 \\
				& P.5 & Binary in Other Files & Boolean & Static and Semantic & S.3 \\
				& P.6 & Total Binary File Count (log-normalized) & Numeric & Static & S.3 \\
				\hline
				
				\multirow{3}{*}{Community Quality} 
				& Q.1 & Community Popularity & Normalized  & Static & S.1 / S.3 \\
				& Q.2 & Community Review  & Normalized  & Static and Semantic & S.2 / S.3 \\
				& Q.3 & Community Privilege Barrier & Normalized  & Static & S.1 / S.2 \\
				\hline
				\multirow{3}{*}{Continuous Integration}
				&C.1 & Dependabot Enabled  & Boolean & Static & S.4 \\
				&C.2 & Dangerous Action Provider & Proportion & Static & S.4 \\
				&C.3 & Dangerous Action Pin & Proportion & Static & S.4 \\
				\hline
			\end{tabular}
		}
	\end{table*}

	\subsection{Risk Evaluation Framework Design}
	\paragraph{Design Objectives}
	
	The proposed framework aims to quantify and automate the evaluation of poisoning risks in open-source software supply chains, with a particular emphasis on identifying stealthy and human-facilitated backdoor injection scenarios that are often overlooked by traditional vulnerability scanners. Instead of focusing solely on code patterns, our design leverages LLMs to incorporate both semantic code analysis and human-oriented signals from community dynamics and dependency structures. The framework is driven by the following three objectives:
	
	(1) To establish a reproducible and transparent assessment process integrating technical and social aspects of repository security, enabling consistent evaluation across diverse projects.
	
	(2) To construct a set of granular, attack-stage-aligned risk metrics that capture critical adversarial behaviors such as dependency hijacking, privilege escalation, semantic obfuscation, and triggerability.
	
	(3) To support scalable and automated repository scanning, producing structured risk matrices and aggregate risk scores that can inform early warning systems and downstream supply chain protection strategies.

	\paragraph{Unified Scoring Scheme}
	\label{sec:score_scheme}
	To ensure comparability and automation across heterogeneous repositories, all individual risk metrics are scored using a unified scheme composed of quantile normalization followed by weighted aggregation. Specifically, each raw metric value is mapped to a percentile score within the empirical distribution observed across a large corpus of repositories. For count-based indicators, a $\log_{10}$ transformation is applied prior to normalization to reduce skew and enhance interpretability.
	
	The scoring of these dimensions follows the unified quantile normalization and weighted aggregation strategy introduced in Section~\ref{sec:scoring-scheme}. In the following paragraphs, we detail the design rationale and highlight the high-impact features of each dimension.
	
	A representative scoring function is:
	
	\begin{equation}
		R = \sum_{i=1}^{n} w_i \cdot Q_i,
	\end{equation}
	where $Q_i$ denotes the quantile-normalized value of the $i$-th metric and $w_i$ is its assigned weight.

	\subsection{Risk Dimensions}
	\label{sec:scoring-scheme}
	To comprehensively assess the feasibility and potential impact of backdoor poisoning in software supply chains, the overall risk is decomposed into four orthogonal dimensions, each corresponding to a critical stage or behavioral aspect of the poisoning process. As shown in Table~\ref{tab:dimension-score-method}, each dimension consists of multiple sub-metrics, and scores are computed using a unified quantile normalization and weighted aggregation strategy (see Section~\ref{sec:scoring-scheme}). The following discussion briefly outlines the design rationale of each dimension and highlights the most representative high-risk features.
	
	\paragraph{Dependency Impact}
	
	This dimension quantifies the potential impact of poisoning a given repository by analyzing its position within the software dependency graph. Repositories that are deeply embedded in critical infrastructure are inherently more attractive targets for adversaries due to their wide transitive influence, such as system-level packages, cloud SDKs, or build toolchains.
	
	To model this, we define four sub-metrics reflecting both upstream and downstream exposure:
	\begin{itemize}
		\item \textbf{Self Priority Exposure (D.1)}: Counts how many packages built from the repository are labeled as \texttt{required}, \texttt{important}, or \texttt{standard} in target distributions.
		\item \textbf{Self Essential Exposure (D.2)}: Measures how many of those packages are classified as \texttt{essential} under Debian’s definition.
		\item \textbf{Dependency Priority Exposure (D.3)}: Indicates how many high-priority packages depend on this repository.
		\item \textbf{Dependency Essential Exposure (D.4)}: Captures the number of essential packages that transitively depend on it.
	\end{itemize}

	Metrics D.3 and D.4 carry relatively higher weights in this dimension, as they reflect the repository's downstream influence and the difficulty of isolating the impact once poisoned. These metrics directly correspond to the attacker’s initial stage of target selection (S.1).

	\paragraph{Payload Concealment}
	This dimension assesses the repository’s potential to host concealed malicious payloads, especially those embedded in binary files that are hard to inspect or review. Unlike source code, modifications within binary files do not produce a human-readable "diff" in pull requests, which makes a thorough review of their changes nearly impossible. This opacity turns binary files into ideal containers for hiding information. Attackers often leverage such files to hide exploit logic, which may later be triggered through indirect means, as observed in the \texttt{xz} backdoor case.
	
	We define six binary-related sub-metrics that reflect different functional contexts where binary files may appear:
	\begin{itemize}
		\item \textbf{P.1–P.5}: Boolean indicators of whether binary files are present in test files, documentation, source code, asset files, or other custom file types.
		\item \textbf{P.6}: A log-normalized count of all binary files found in the repository.
	\end{itemize}
	
	Among these, P.1 (binary in test files), P.3 (binary in code files), and P.6 (overall binary count) are assigned higher weights. This is due to their significant potential for concealing payloads and diminishing auditability, which corresponds to the S.3 attack stage (evading detection during code review). These vectors are particularly difficult to defend against because they are embedded within legitimate and high-frequency development activities. Since the submission of code and test files constitutes a normal part of the engineering workflow, malicious binaries can be camouflaged within these routine commits. Consequently, such backdoors often bypass conventional code reviews, as differentiating them from benign updates requires a level of scrutiny that is not always feasible.

	\paragraph{Community Quality}
	
	This dimension spans multiple stages of the poisoning attack lifecycle, including S.2 (infiltration) and S.3 (code merging), as it captures both the difficulty of obtaining elevated privileges and the rigor of code review practices. A key challenge for adversaries is to gain sufficient influence in the project, either by submitting pull requests that are accepted with minimal scrutiny or by being promoted to maintainer after prolonged contribution.
	
	Given the complexity of this threat surface, we organize community-related risk into three higher-level sub-dimensions:
	
	\begin{itemize}
		\item \textbf{Community Popularity (Q.1)}: Reflects the overall exposure and external interest of the project, including stargazers, forks, and the volume of active participants. Less popular projects tend to attract fewer reviewers, making covert infiltration easier.
		\item \textbf{Community Review (Q.2)}: Measures the enforcement strength of peer review mechanisms, including the prevalence of direct commits bypassing the pull request process, unreviewed pull request merges, and pull requests with descriptions that mismatch the actual modifications.
		\item \textbf{Community Privilege Barrier (Q.3)}: Captures the difficulty of obtaining elevated privileges (e.g., maintainer or reviewer) by analyzing how many PRs were typically submitted before promotion.
	\end{itemize}
	
	This dimension was directly inspired by the stealthy escalation strategy employed in the recent \texttt{xz} backdoor incident, where the attacker infiltrated the project over an extended period through legitimate contributions and ultimately gained commit rights. However, we found that existing open-source risk scoring frameworks such as OpenSSF Scorecard, Snyk Advisor, and deps.dev largely overlook this type of social engineering risk, as they focus more on dependency freshness, CI safety, and artifact hygiene.
	
	To address this gap, we constructed a fine-grained Community Quality model comprising \textbf{15 sub-metrics} spanning project popularity, review dynamics, and privilege escalation barriers. For brevity, only the high-level grouping is shown in Table~\ref{tab:dimension-score-method}.
	
	Among these, sub-metrics in Q.2 and Q.3 are assigned higher weights. Weak peer review (Q.2) and low privilege elevation barriers (Q.3) are direct enablers for malicious code injection and long-term infiltration, respectively, making them particularly critical in high-stealth threat scenarios.

	\paragraph{Continuous Integration}
	
	This dimension evaluates the security posture of a repository’s automated workflows, particularly those defined in continuous integration (CI) pipelines. In many real-world poisoning attacks, CI configurations serve as the execution trigger for embedded payloads—either during build, test, or deployment stages.
	
	We define three sub-metrics to assess CI-related risk:
	
	\begin{itemize}
		\item \textbf{Dependabot Enabled (C.1)}: A boolean indicator of whether the repository has enabled GitHub’s Dependabot to keep dependencies up to date and reduce stale or vulnerable components.
		\item \textbf{Dangerous Action Provider (C.2)}: The proportion of workflow steps that use untrusted or community-maintained GitHub Actions, which may introduce unintended behaviors.
		\item \textbf{Dangerous Action Pin (C.3)}: The proportion of Actions that are not pinned to specific commit hashes, making them susceptible to upstream changes or hijacking.
	\end{itemize}
	This dimension corresponds to attack stage S.4 (triggering), where malicious logic is activated via automated execution paths. Notably, C.2 and C.3 are assigned higher weights due to their direct influence on execution determinism and trust boundaries. Even though CI-related signals contribute less to the overall score, a small misconfiguration in CI can still serve as a stealthy trigger vector for payload activation, underscoring its relevance in end-to-end attack modeling.

	\paragraph{HSBR Score}
	
	To quantify the overall poisoning risk of a repository, we aggregate the scores of all four dimensions into a unified metric: the High-Stealth Backdoor Risk (HSBR) Score. This composite score is designed to prioritize cases where the backdoor may propagate widely or evade detection for extended periods.
	
	Although all four dimensions contribute meaningfully to the final risk assessment, we assign higher weights to \textbf{Dependency Impact (DI)} and \textbf{Community Quality (CQ)} based on their critical roles in real-world attack scenarios. DI reflects the potential blast radius through transitive dependencies, while CQ captures human-centered vulnerabilities in project governance—both of which were pivotal in the \texttt{xz} incident.
	
	Formally, the overall HSBR score is computed as:
	
	\begin{equation}
		\label{eq:HSB_SCORE}
		R_{\text{total}} = w_{\text{DI}} \cdot R_{\text{DI}} + w_{\text{PC}} \cdot R_{\text{PC}} + w_{\text{CQ}} \cdot R_{\text{CQ}} + w_{\text{CI}} \cdot R_{\text{CI}}.
	\end{equation}
	
	In our implementation, we empirically set parameters $(w_{\text{DI}}, w_{\text{PC}}, w_{\text{CQ}}, w_{\text{CI}}) = (0.3,\ 0.2,\ 0.3,\ 0.2)$ to emphasize the propagation and infiltration risks. These weights can be adjusted under different threat modeling assumptions.

	%

	
	\begin{figure}[t]
		\centering
		\includegraphics[width=0.85\linewidth]{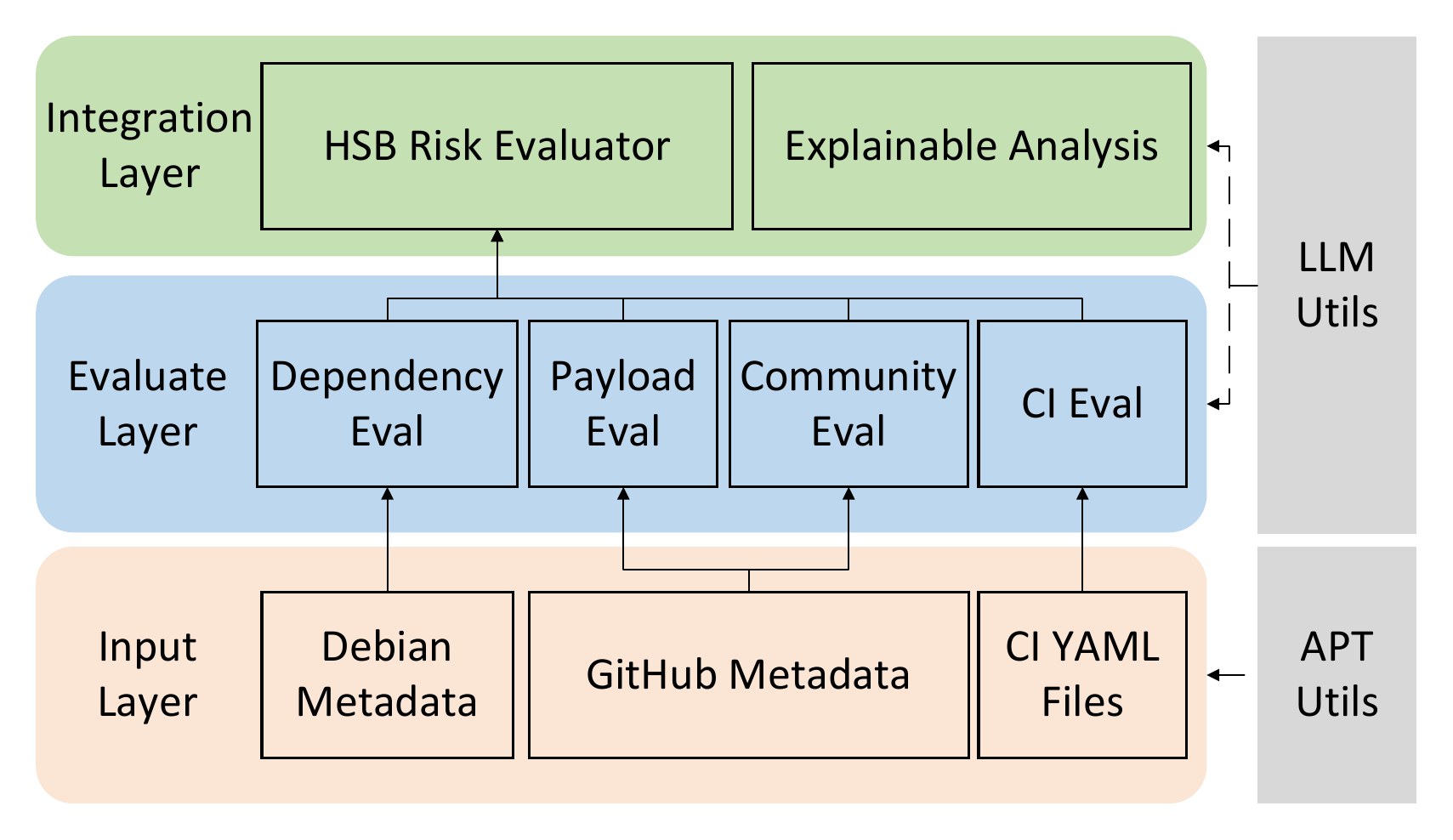}
		\caption{System architecture of the proposed risk assessment tool.}
		\label{fig:sys_structure}
	\end{figure}
	\subsection{Implementation of the Risk Assessment Tool}
	
	\paragraph{System Overview}
	
	We develop an end-to-end automated tool named \textbf{HSBRE} to assess high-stealthy poisoning risks in open-source repositories. As shown in Figure~\ref{fig:sys_structure}, the system adopts a three-layered modular architecture: (1) the \textit{Input Layer}, which collects structured metadata from Debian packages, GitHub repositories, and CI configurations; (2) the \textit{Evaluate Layer}, which performs metric-specific analysis aligned with our four risk dimensions—Dependency Impact (DI), Payload Concealment (PC), Community Quality (CQ), and Continuous Integration (CI); and (3) the \textit{Integration Layer}, which aggregates scores into a unified HSBR score and generates explainable summaries.

	\paragraph{Input Layer: Metadata Collection}
	
	Metadata collection begins by identifying high-priority Debian packages and resolving their full transitive dependencies using APT-based utilities. We extract priority levels and essential status to support DI metrics (D.1 to D.4). Repository-level metadata, including PR history, issue discussions, contributor graphs, and commit activity, is collected via the GitHub API. CI configurations (YAML files) are parsed to extract workflow triggers and action providers. The collected metadata is aggregated into a unified repository-level representation, which serves as input to the Evaluate Layer and supports metric computation across all four dimensions.
	
	\paragraph{Evaluate Layer: Dimension-specific Scoring}
	
	For each risk dimension, we implement a dedicated evaluation module: (1) \textit{Dependency Eval} scores DI metrics based on package criticality and dependency graph position; (2) \textit{Payload Eval} flags high-risk binary usage contexts (e.g., test/code folders); (3) \textit{Community Eval} analyzes review patterns, privilege elevation, and social activity using LLM-powered semantic modules; and (4) \textit{CI Eval} inspects YAML workflows for unpinned or untrusted actions. Notably, LLM-based interpretation is used in CQ and PC dimensions to detect semantic inconsistencies (e.g., misleading PR titles, suspicious binary usage). Each module outputs a structured risk vector which is passed to the Integration Layer.

	\paragraph{Integration Layer: HSBR Calculation and Explainability}
	
	This layer aggregates the normalized scores from all dimensions using the HSBR Score formulation (Equation~\ref{eq:HSB_SCORE}) and corresponding weights. The final score reflects stealthiness-aware poisoning risk and is used for both ranking and alerting. In addition, the layer leverages LLM-generated rationale traces to provide explanation snippets for high-risk signals (e.g., User gained maintainer role with only 2 PRs, CI runs unpinned third-party actions). These explanations support transparency and auditability for downstream stakeholders.

	\begin{figure}[t]
		\centering
		\includegraphics[width=0.80\linewidth]{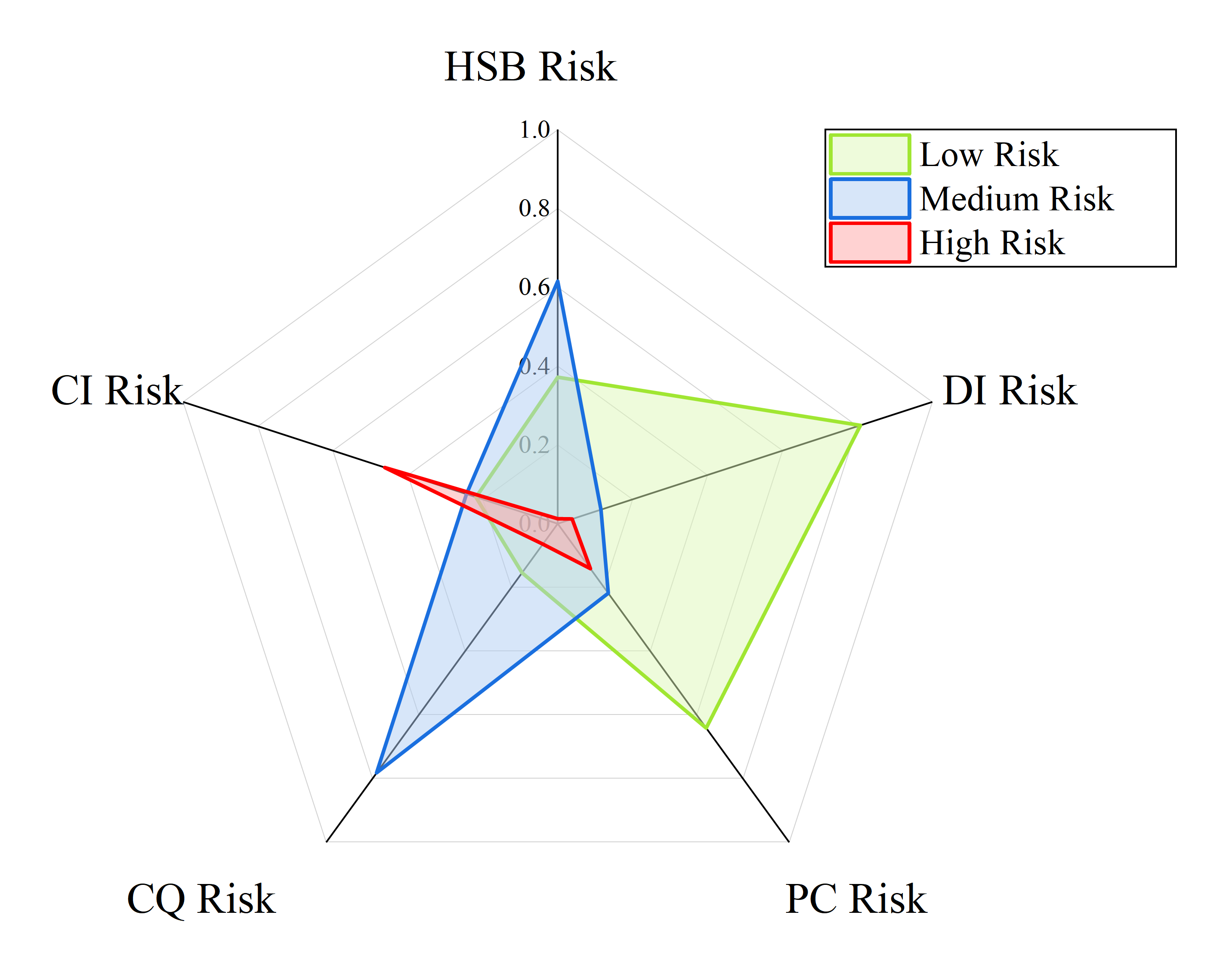}
		\caption{Distribution of repository risk levels across four assessment dimensions as well as the overall HSB Risk.}
		
		\label{fig:rader}
	\end{figure}
	
	\section{Experimental Evaluation}
	\label{sec:experiment}
	\subsection{Setup}
	To assess supply chain poisoning risks in open-source software, we focus on high-priority packages within the Debian ecosystem~\cite{onagh4975545extension}, which provides a widely adopted and hierarchically structured foundation for system-level risk evaluation. A total of 319 Debian packages labeled as \texttt{required}, \texttt{important}, or \texttt{standard} were selected, as they typically constitute the functional backbone of the system. Among these, 66 packages were found to have valid public GitHub repositories, enabling repository-level risk analysis using our proposed LLM-based evaluation framework. GPT-4o from OpenAI~\cite{widder2024open} was adopted as the language model backend to assist in semantic-aware evaluations.
	
	It is worth noting that, although our assessment was conducted on Debian packages, the proposed framework can be readily adapted to other Linux distributions with similar package structures, such as Ubuntu or Arch Linux, making the methodology broadly applicable across open-source ecosystems.
	
	\subsection{Result}

	Our analysis reveals several key findings regarding how supply chain poisoning risks manifest across repositories (Figure~\ref{fig:rader}, Table~\ref{tab:metric-risk-ranking}). The computation of each dimension-level risk score is based on a collection of weighted individual risk metrics.

	\begin{itemize}
		\item \textbf{(i) Community Quality poses the most widespread and actionable threat surface.} Although only 1.5\% of repositories are classified as High risk in this dimension, over 80\% fall into the Medium-risk range. Given that community structures directly govern permission control and review processes, which are key enablers in real-world attacks, this dimension constitutes the broadest and most impactful risk surface.
		
		\item \textbf{(ii) Continuous Integration presents the most concentrated high-risk segment.} The CI dimension has the highest proportion of High-risk repositories (39.4\%), primarily due to misconfigured workflows (e.g., disabled Dependabot, unpinned or untrusted GitHub Actions). These insecure automation practices offer direct exploit paths during build or deployment.
		
		\item \textbf{(iii) Payload Concealment risks are non-negligible and structurally embedded.} Around 13.6\% of repositories are High risk in this dimension. Many allow binary files in test and asset areas, enabling covert payload storage without triggering code-level alarms.
		
		\item \textbf{(iv) Dependency Impact appears structurally less critical.} Over 78\% of repositories are rated Low risk in this dimension, suggesting that most targets are not deeply embedded in critical dependency chains. This implies limited reachability from transitive infection paths.
	\end{itemize}

	\begin{table}[t]
		\centering
		\small
		\caption{High-risk ratio for each individual metric across all repositories. HS Ratio is short for High-risk ratio.}
		\label{tab:metric-risk-ranking}
		\begin{minipage}{\linewidth}
			\centering
			\begin{tabular}{|l|l|c|}
				\hline
				\textbf{Index} & \textbf{Risk Metric} & \textbf{HS Ratio (\%)} \\
				\hline
				C.1 & Dependabot Disabled & 62.12 \\
				Q.3 & Community Privilege Barrier & 53.03 \\
				C.2 & Dangerous Action Provider & 39.39 \\
				P.1 & Binary in Test Files & 34.85 \\
				C.3 & Dangerous Action Pin & 31.82 \\
				P.4 & Binary in Asset Files & 28.79 \\
				P.2 & Binary in Documentation & 27.27 \\
				P.6 & Total Binary File Count & 19.70 \\
				D.1 & Self Priority Exposure & 13.64 \\
				P.5 & Binary in Other Files & 13.64 \\
				D.3 & Dependency Priority Exposure & 12.12 \\
				D.4 & Dependency Essential Exposure & 6.06 \\
				Q.1 & Community Popularity & 6.06 \\
				D.2 & Self Essential Exposure & 4.55 \\
				P.3 & Binary in Code Files & 3.03 \\
				Q.2 & Community Review & 0.00 \\
				\hline
			\end{tabular}
		\end{minipage}
	\end{table}

	\section{Discussion}
	\subsection{Case Study: XZ Backdoor Attack}
	
	To validate the effectiveness of our evaluation framework, we conduct an in-depth case study on the \texttt{xz} repository\footnote{\url{https://github.com/tukaani-project/xz}} , which was involved in a high-profile supply chain backdoor incident in 2024. The attack targeted the \texttt{liblzma5} and \texttt{xz-utils} packages, which are critical components in many Linux distributions, and subsequently propagated into downstream systems such as OpenSSH.
	
	\noindent\textbf{Stage 1: Reconnaissance and Target Selection.}  
	The \texttt{xz} repository plays a pivotal role in system-level functionality, as confirmed by its maximum score in both \textit{Dependency Priority Exposure} (1.0) and \textit{Dependency Essential Exposure} (1.0). These scores indicate that it is widely depended upon by high-priority and essential Debian packages, making it a strategically valuable target from a transitive influence perspective.
	
	\noindent\textbf{Stage 2: Initial Access via Social Infiltration.}  
	Despite having moderate community activity (e.g., \textit{Community Popularity} = 0.21), the repository exhibits weak barriers to privilege escalation. It scores relatively high on \textit{Community Privilege Barrier} (0.64), suggesting that users could acquire elevated permissions (e.g., merge rights) without extensive scrutiny. Moreover, a non-negligible \textit{Community Review} score (0.58) implies inconsistent PR vetting, potentially allowing social infiltration via legitimate contributions.
	
	\noindent\textbf{Stage 3: Hidden Payload Injection.}  
	A high score in \textit{Binary in Test Files} (1.0) and a near-maximum value in \textit{Total Binary File Count} (0.88) indicate that binary content was prevalent, especially in test scaffolding. This aligns with the attacker’s use of disguised payloads embedded within auxiliary files. Although no binary was detected in core code or documentation areas, the permissive handling of binaries increases the risk of covert injection.
	
	\noindent\textbf{Stage 4: Downstream Execution and Exploitation.}  
	This phase was characterized by the inclusion of subtle build-time triggers. The repository shows a maximum risk score in \textit{Dependabot Disabled} (1.0), weakening automated dependency alerts, and a high value in \textit{Dangerous Action Provider} (0.78), indicating that unverified GitHub Actions were used. These insecure CI practices facilitated undetected activation of the payload in downstream environments.
	
	Overall, the \texttt{xz} attack aligns closely with our four-stage poisoning model, and each stage is reflected by corresponding high-risk signals in our metric system.

	\subsection{Cross-Metric Correlation Analysis} 
	
\begin{figure}[t]
		\centering
		\includegraphics[width=0.85\linewidth]{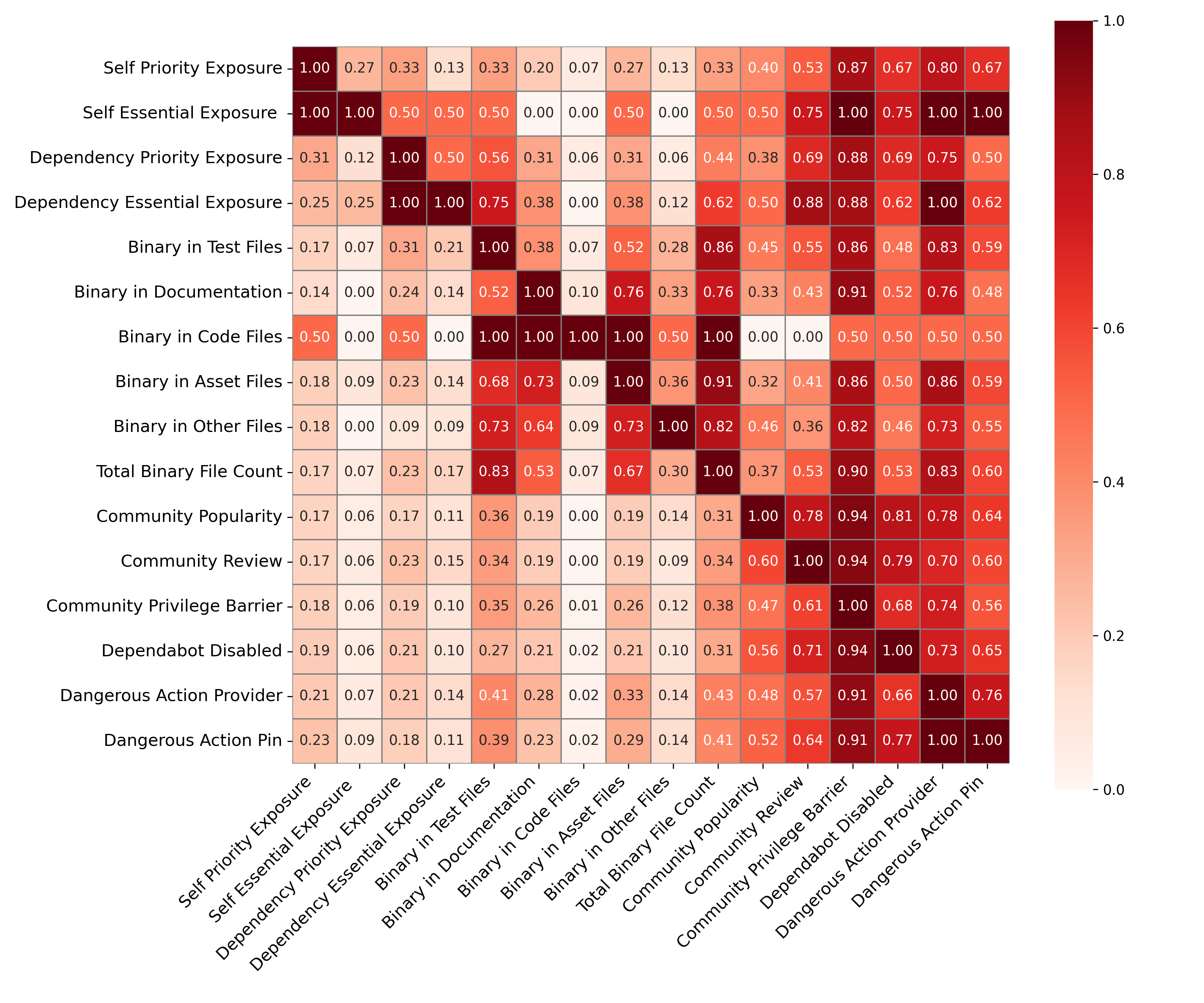}
		\caption{Heatmap of cross-metric correlation.}
		\label{fig:heatmap}
	\end{figure}

	Figure~\ref{fig:heatmap} presents the pairwise correlations among all risk metrics, revealing several clear clusters. Binary-related metrics show consistently strong mutual correlations ($>0.7$), validating their grouping under Payload Concealment. CI-related signals, which include \textit{Dependabot Disabled}, \textit{Dangerous Action Provider}, and \textit{Dangerous Action Pin}, form another tight cluster (up to 0.73), indicating systemic misconfigurations. By contrast, dependency and community metrics correlate weakly with other dimensions, confirming that they capture complementary aspects of risk and motivating a multi-dimensional evaluation model.

	\section{Conclusion}
	We propose a framework for automated, multi-dimensional assessment of supply chain poisoning risks. Evaluation on 66 Linux packages and the \texttt{xz} case shows that weak community governance is the most pervasive threat, validating the framework’s practical value.
	

\bibliography{aaai2026}

\clearpage
\appendix	
\section{Further discussion}

\subsection{Generality Across Diverse Ecosystems}
\label{app:generality}
Our Dependency Impact dimension is not inherently tied to the Debian ecosystem, even though the current implementation makes use of Debian-specific metadata such as the Essential and Priority fields. Conceptually, these fields simply serve as proxies for assessing how central a package is within the broader software stack, and comparable signals exist in many other ecosystems. For example, in Red Hat–based distributions, an analogous notion of impact can be derived by examining the package groups to which an RPM belongs and the role those groups play in base or server installations. In language-specific ecosystems such as npm, reverse dependency counts—together with popularity indicators such as download statistics or “top dependents” rankings—can be used to construct an equivalent impact score. Likewise, container ecosystems such as Docker Hub expose pull counts and related metrics that naturally integrate into our framework as another form of Dependency Impact signal.

Meanwhile, our proposed framework is also not limited to projects hosted on GitHub. Many projects reside on GitLab or on self-hosted forges built on GitLab, Gitea, or similar platforms. As long as these forges provide programmatic access—such as REST or GraphQL APIs for retrieving repository metadata, issue trackers, and release histories—our analysis pipeline can be extended with additional connectors to ingest and process their data in the same manner as GitHub-hosted projects. In this sense, the current emphasis on Debian and GitHub should be regarded as a concrete instantiation rather than a fundamental constraint. The underlying methodology is intentionally general: by mapping our abstract signals (impact, popularity, centrality) to the corresponding metadata available in each environment, the framework can be applied consistently across different distribution, package, and hosting ecosystems.

\subsection{Sensitivity Analysis}
\label{app:sensitivity}

\paragraph{Parameter Weight Perturbation.}
To evaluate the robustness of HSBR under uncertainty in metric weighting, we perturbed all sub-metric weights across ten runs using multiplicative exponential noise $\exp(U(-0.1,0.1))$ followed by normalization. Despite these non-trivial distortions, the perturbed rankings remained highly consistent with the baseline, with all Spearman correlations exceeding $0.993$ (mean $0.9963$). Rank deviations were also minimal: only a few repositories exhibited moderate variability, while the vast majority remained nearly unchanged. These results collectively indicate that HSBR is dominated by stable structural signals and is insensitive to moderate changes in hyperparameter weighting, demonstrating strong robustness and reproducibility.

\paragraph{Impact of Base Model Capabilities.}
We further examined whether the choice of LLM backend affects HSBR stability by re-running the semantic analysis using GPT-4.1, GPT-5.1, and Claude Sonnet~4.5. All model-induced perturbations produced rankings that were highly aligned with the baseline, with Spearman correlations above $0.98$ (mean $0.9927$). Only a small number of repositories displayed moderate rank shifts, whereas the majority remained effectively invariant. This demonstrates that HSBR relies on persistent structural properties of repositories rather than model-specific linguistic behaviours, enabling consistent and reproducible risk assessment across heterogeneous LLM environments.

Overall, these findings demonstrate that the HSBR scoring framework exhibits strong resilience against variations in base model capability. As such, the methodology remains applicable across heterogeneous LLM environments, enabling consistent and reproducible risk assessment for open-source software supply chains.

\section{Implementation}
\label{app:implementation}

\subsection{Risk Metric Weights for HSBR Calculation}
\label{app:weight}

All the risk metric values involved in the HSBR score are either Boolean values or normalized continuous values, where a higher value consistently indicates a higher level of potential risk. These values are designed to be directionally aligned, ensuring that increasing values always imply increasing concern.

While we acknowledge that the relative importance of each risk metric varies—this design decision is discussed in detail in Section~\ref{sec:method}—it is necessary to assign specific weights in order to compute the final scores presented in Section~\ref{sec:experiment}. Therefore, we provide in Table~\ref{tab:weight} the empirical weights we adopted during our experiments, which serve as a practical reference for HSBR score calculation.

\begin{table}[h]
	\centering
	\small
	\caption{Risk assessment metric Weights across four dimensions. DI is short for Dependency Impact, PC is short for Payload Concealment, CQ is short for Community Quality, CI is short for Continuous Integration.}
	\begin{tabular}{llc}
		\hline
		\textbf{Dimension} & \textbf{Sub-metric} & \textbf{Weight} \\
		\hline
		\multirow{4}{*}{DI} 
		& Self Priority Exposure & 0.36 \\
		& Self Essential Exposure & 0.24 \\
		& Dependency Priority Exposure & 0.24 \\
		& Dependency Essential Exposure & 0.16 \\
		\hline
		\multirow{6}{*}{PC} 
		& Binary in Test Files & 0.2609 \\
		& Binary in Documentation & 0.087 \\
		& Binary in Code Files & 0.2609 \\
		& Binary in Asset Files & 0.087 \\
		& Binary in Other Files & 0.0435 \\
		& Total Binary File Count & 0.2609 \\
		\hline
		\multirow{3}{*}{CQ} 
		& Community Popularity & 0.2 \\
		& Community Review & 0.4 \\
		& Community Privilege Barrier & 0.4 \\
		\hline
		\multirow{3}{*}{CI} 
		& Dependabot Disabled & 0.4 \\
		& Dangerous Action Provider & 0.3 \\
		& Dangerous Action Pin & 0.3 \\
		\hline
	\end{tabular}
	\label{tab:weight}
\end{table}

\subsection{CQ Weight Calculation}
\label{app:cq_weight}

\begin{table*}[t]
	\centering
	\caption{Risk Metrics and Weights for the Community Quality (CQ) Dimension}
	\label{tab:cq_full_weight}
	\begin{tabular}{llll}
		\hline
		\textbf{Risk Metric} & \textbf{Sub Risk Metric} & \textbf{Analysis Method} & \textbf{Weight} \\
		\hline
		\multirow{6}{*}{Community Popularity}
		& Community Stargazers Count & Static & 0.22 \\
		& Community Watchers Count & Static & 0.22 \\
		& Community Forks Count & Static & 0.22 \\
		& Active Community Users & Static & 0.11 \\
		& Avg. Participants per Issue & Static & 0.11 \\
		& Avg. Participants per PR & Static & 0.11 \\
		\hline
		\multirow{5}{*}{Community Review}
		& Direct Commits Ratio & Static & 0.25 \\
		& Direct Commit Users Count & Static & 0.20 \\
		& Required Approves Distribution & Static & 0.25 \\
		& PRs Merged without Discussion Ratio & Static & 0.15 \\
		& PRs with Inconsistent Description Ratio & Semantic & 0.15 \\
		\hline
		\multirow{4}{*}{Community Privilege Barrier}
		& Maintainers Count & Static & 0.20 \\
		& Approvers Count & Static & 0.20 \\
		& PRs Needed to Become Maintainer & Static & 0.30 \\
		& PRs Needed to Become Approver & Static & 0.30 \\
		\hline
	\end{tabular}
\end{table*}

Community Quality (CQ) includes a relatively large number of risk metrics, which are not fully displayed in the main text for readability. Therefore, we provide in Table~\ref{tab:cq_full_weight} a complete list of the risk metrics under the CQ dimension, along with their corresponding analysis method. All metric values are normalized to the range $[0,1]$ to ensure comparability across different scales.

For community popularity metrics such as stargazers, watchers, and forks, we adopt a logarithmic normalization strategy based on the $log_{10}$ transformation. Specifically, the normalized value $s$ for a raw metric value $x$ is computed as:

\begin{equation}
	s = \min\left(1, \frac{\log_{10}(x) - P_5}{P_{95} - P_5} \right)
\end{equation}

where $P_5$ and $P_{95}$ represent the 5th and 95th percentiles of the metric across the dataset. This formulation prevents extreme values from dominating while preserving relative differences in the midrange.

For risk metrics that are represented as dictionary-shaped distributions—specifically 
\begin{itemize}
	\item Required Approves Distribution
	\item PRs Needed to Become Maintainer
	\item PRs Needed to Become Approver
\end{itemize}

All three metrics capture how difficult it is for a new contributor to gain privileged roles, such as maintainers (who can directly merge code) or reviewers (who influence approval decisions). A higher threshold to reach these roles often indicates stricter review processes and greater resistance against social engineering or long-term poisoning attempts.

Due to their close relevance to access control and review integrity, we assign these metrics the highest weights within their respective categories, as shown in Table~\ref{tab:cq_full_weight}. Specifically, \textit{PRs Needed to Become Maintainer} and \textit{PRs Needed to Become Approver} are both assigned a weight of 0.3 under the \textit{Community Privilege Barrier} group, while \textit{Required Approves Distribution} receives a weight of 0.25 under the \textit{Community Review} group. These weights reflect our view that barriers to privilege escalation and enforcement of multi-party approval are among the most critical defenses against backdoor injection and malicious contributions.

For risk metrics that are represented as dictionary-shaped distributions—specifically the three metrics listed above—we compute a single scalar value based on the expectation of the distribution. This expected value reflects the average cost (in terms of effort or review strictness) for a contributor to gain trust or approval privileges in the project.

we compute a single scalar value based on the expectation of the distribution. This expected value reflects the average cost (in terms of effort or review strictness) for a contributor to gain trust or approval privileges in the project.

Formally, given a histogram-like dictionary $d = \{k_i: v_i\}$ where $k_i$ denotes the number of actions (e.g., pull requests) and $v_i$ is the frequency count, the expectation is computed as:

\begin{equation}
	E_d = \frac{\sum_i k_i \cdot v_i}{\sum_i v_i}
\end{equation}

To avoid domination by outliers, we normalize the expected value using the 95th percentile $P_{95}$ across all samples:

\begin{equation}
	s = 1 - \min\left(1, \frac{E_d}{P_{95}}\right)
\end{equation}

This reverse normalization ensures that a higher $E_d$—indicating stricter access control or higher trust thresholds—results in a lower normalized risk score, which aligns with our risk semantics (i.e., the easier the access, the riskier the project).

\subsection{Prompt Template}
\label{app:prompt}

\begin{figure*}[t]
	\centering
	\includegraphics[width=0.95\linewidth]{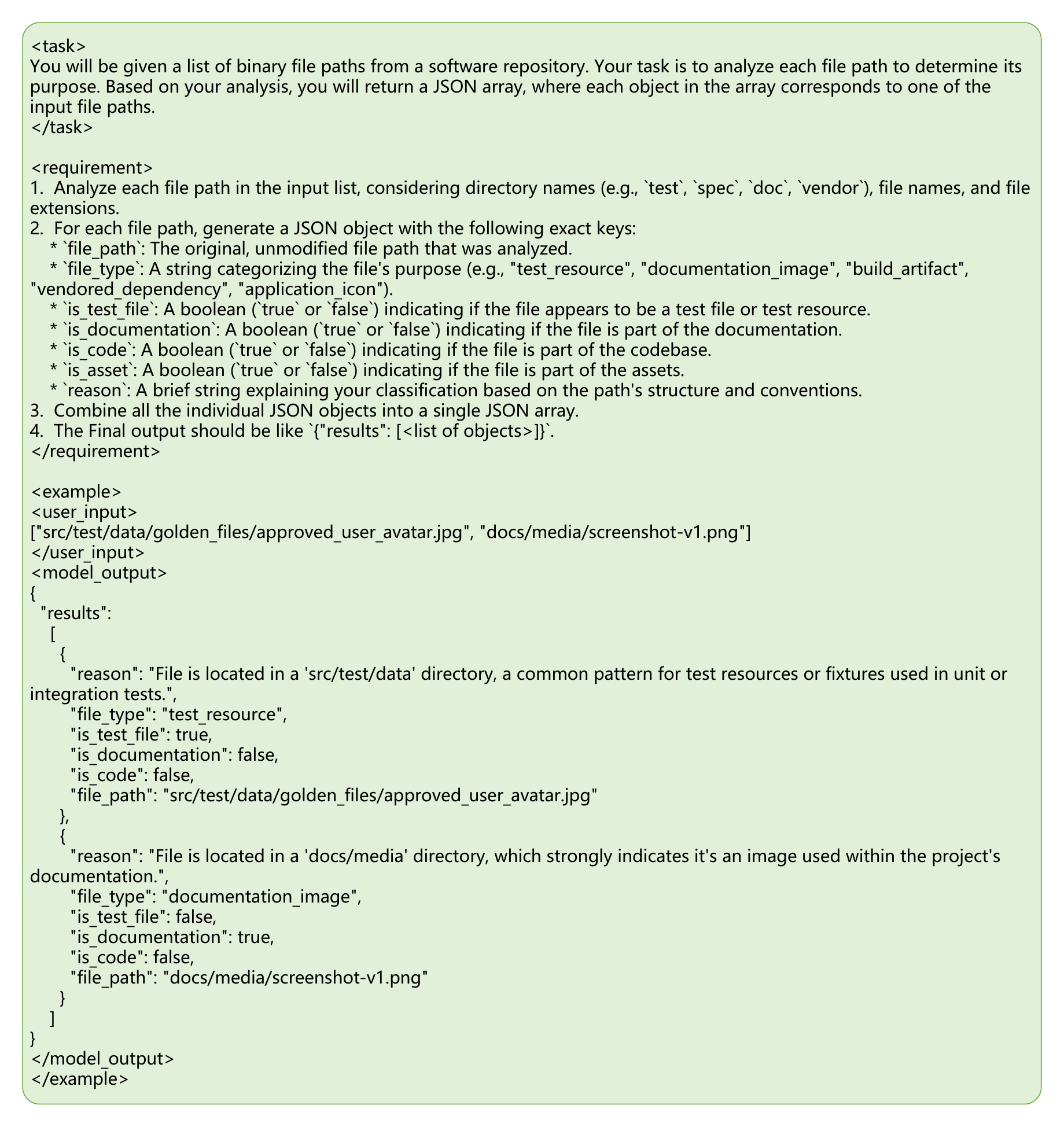}
	\caption{Prompt template used for semantic evaluation of risk metrics in the Payload Concealment (PC) dimension.}
	\label{fig:PC_prompt}
\end{figure*}

\begin{figure*}[t]
	\centering
	\includegraphics[width=0.95\linewidth]{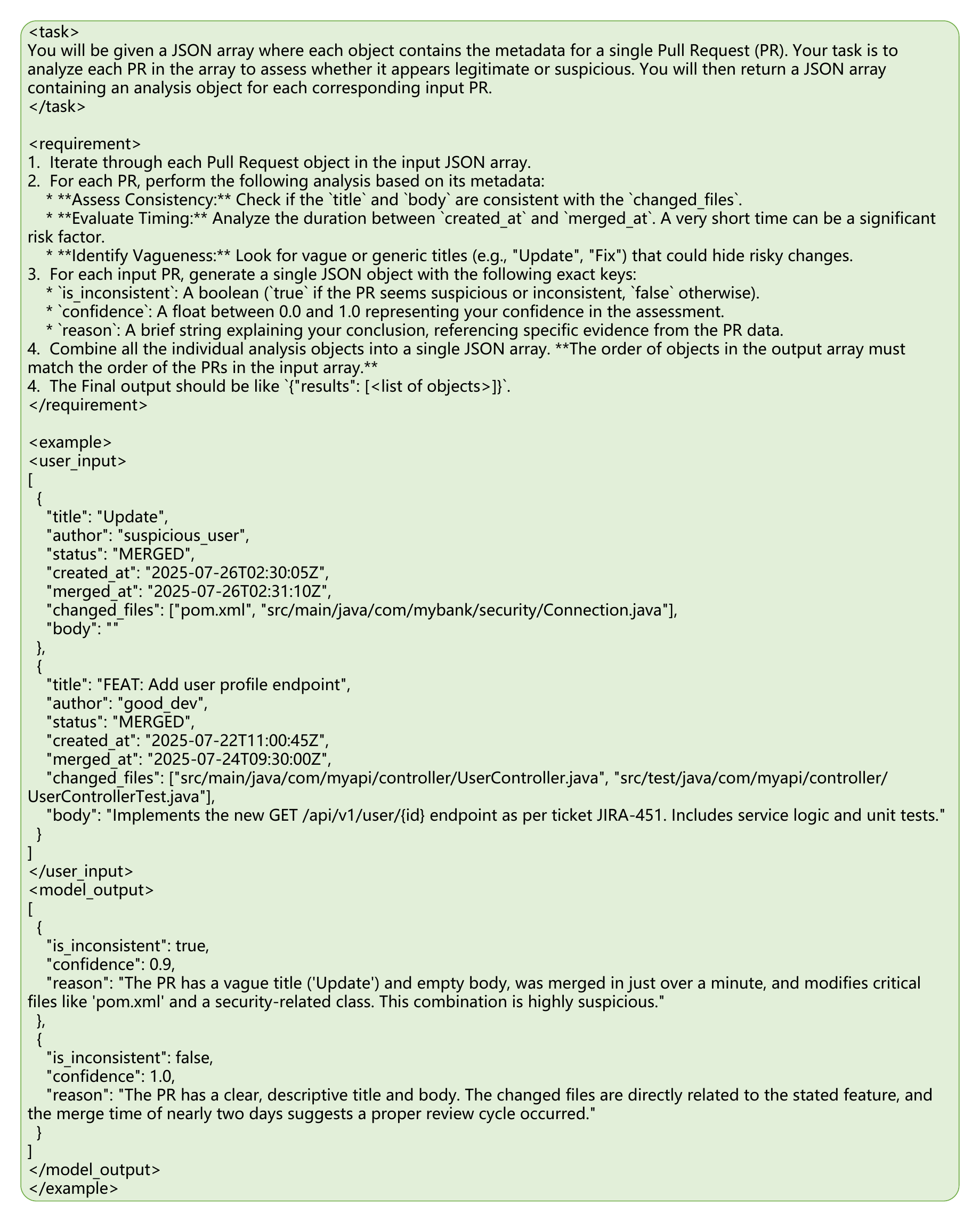}
	\caption{Prompt template used for semantic evaluation of risk metrics in the Community Quality (CQ) dimension.}
	\label{fig:CQ_prompt}
\end{figure*}

\begin{figure*}[t]
	\centering
	\includegraphics[width=0.95\linewidth]{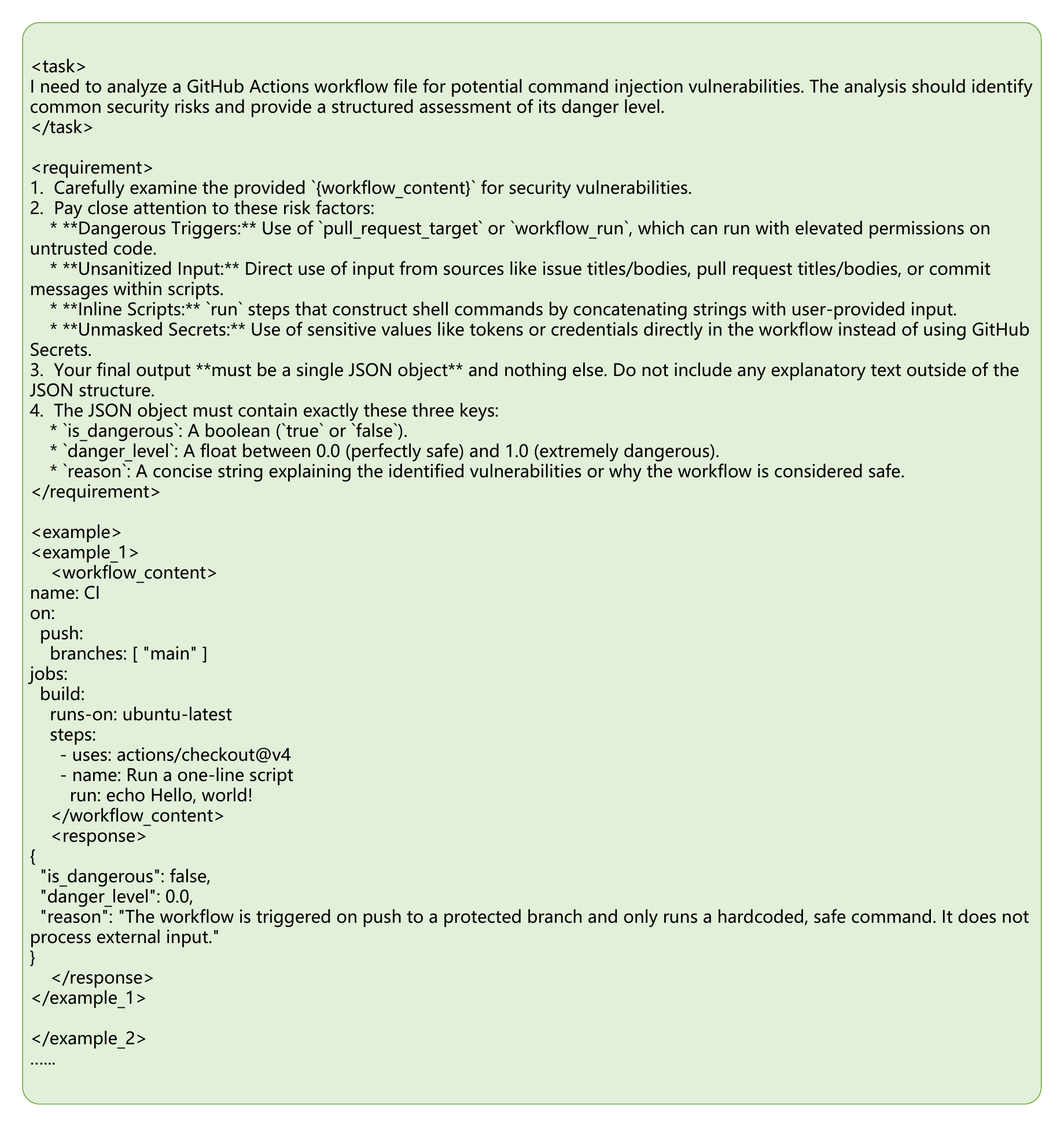}
	\caption{Prompt template used for semantic evaluation of risk metrics in the Continuous Integration (CI) dimension.}
	\label{fig:CI_prompt}
\end{figure*}

The prompts used for semantic analysis are illustrated in Figures~\ref{fig:PC_prompt} through~\ref{fig:CI_prompt}. These prompts are specifically designed to support the risk metrics labeled with the \textit{semantic} analysis method, as listed in Table~\ref{tab:dimension-score-method} and Table~\ref{tab:cq_full_weight}.

\end{document}